\documentclass[
]{ceurart}

\usepackage{listings}
\usepackage{csquotes}
\usepackage{xcolor}
\usepackage{natbib}
\usepackage{flushend}
\usepackage{comment}

\newcommand{\todo}[1]{{\color{red}TODO: #1}}
\newcommand{\simo}[1]{{\color{olive}SL: #1}}

\newcommand{\anna}[1]{{\color{purple}Anna: #1}}
\newcommand{\christian}[1]{{\color{blue}CG: #1}}

\begin{document} 
\copyrightyear{2022}
\copyrightclause{Copyright for this paper by its authors.
  Use permitted under Creative Commons License Attribution 4.0
  International (CC BY 4.0).}

\conference{ICCC’22 Workshop: The Role of Embodiment in the Perception of Human \& Artificial Creativity, June 27–28, 2022, Bozen, Italy}

\title{How Does Embodiment Affect the Human\\Perception of Computational Creativity?\\An Experimental Study Framework}

\author[1]{Simo Linkola}[%
email=simo.linkola@helsinki.fi,
]
\cormark[1]
\fnmark[1]

\author[2,3]{Christian Guckelsberger}[%
email=christian.guckelsberger@aalto.fi,
]
\fnmark[1]

\author[1]{Tomi Männistö}[%
email=tomi.mannisto@helsinki.fi,
]

\author[1]{Anna Kantosalo}[%
email=anna.kantosalo@helsinki.fi,
]

\address[1]{Department of Computer Science, University of Helsinki, Helsinki, Finland}
\address[2]{Department of Computer Science, Aalto University, Espoo, Finland}
\address[3]{School of Electronic Engineering and Computer Science, Queen Mary University of London, London, UK}

\cortext[1]{Corresponding author.}
\fntext[1]{Contributed equally and share first authorship}

\maketitle

\begin{abstract}
Which factors influence the human assessment of creativity exhibited by a computational system is a core question of computational creativity (CC) research. Recently, the system's embodiment has been put forward as such a factor, but empirical studies of its effect are lacking. To this end, we propose an experimental framework which isolates the effect of embodiment on the perception of creativity from its effect on creativity per se.~We manipulate not only the system's embodiment but also the human perception of creativity, which we factorise into the assessment of creativity, and the perceptual evidence that feeds into that assessment. We motivate the core framework with embodiment and perceptual evidence as independent and the creative process as a controlled variable, and we provide recommendations on measuring the assessment of creativity as a dependent variable. We propose three types of perceptual evidence with respect to the creative system, the creative process and the creative artefact, borrowing from the popular four perspectives on creativity. We hope the framework will inspire and guide others to study the human perception of embodied CC in a principled manner.
\end{abstract}


\section{Introduction}
Which factors contribute to people's \emph{perception of creativity} in artificial systems has been recognised as a central question in computational creativity (CC) research \citep{colton2008creative_tripod}. Insights into this question can not only support the fairer comparison of existing and the design of new systems but also inform strategies to foster people's acceptance of AI systems as creative contributors.

By \enquote{the human perception of creativity}, we denote a person's subjective assessment of an AI system's creativity based on available evidence. Some of the factors identified or hypothesised to affect this perception -- assuming that the creative process and product remain constant -- are the human assessor's competence \citep{lamb2015} and technological optimism \citep{mumford15}, the system's motivational mechanism \citep{guckelsberger2017addressing} and the framing of its creative process \citep{charnley2012notion}, including the description of \enquote{life experiences}~\citep{Colton2020}. In this paper, we focus on a factor that has only recently been hypothesised to affect the human perception of creativity \citep{survey2021embodiment}
: the system's \emph{embodiment} as -- roughly speaking -- the shape of its body and its means to interact with the environment \citep[cf.][]{ziemke2003s}.

A recent survey of publications~\citep{survey2021embodiment} at the International Conference on Computational Creativity (ICCC) attests a long-lasting, steady and recently increasing interest in engineering autonomous and co-creative, embodied CC systems in various domains.  
Crucially though, the survey has not found a single study offering generalising, empirical insights about the effect of a system's embodiment on the perception of creativity. This is highly problematic in that such an effect could be substantial; some even hypothesise that, given a large \enquote{embodiment gap} between a system and its observer, the system's creativity might remain entirely unnoticed \citep{guckelsberger2017addressing}. 

We pave the way towards gaining such insights by proposing a framework for conducting empirical studies on how a system's embodiment impacts the human perception of its creativity. The framework is intended for and relies on comparing this facet of human perception between two or more systems. Within it, we highlight the importance and discuss the potential values and relationship of two independent variables: the \emph{participant's perceptual evidence}, and the \emph{system's embodiment}. By manipulating these variables, we can study what kind of observations about an artificial system lead to the attribution of creativity, and how. We support our discussion with examples but focus on the domain of \emph{visual art}.

Crucially, we target the effect of embodiment on the \emph{perception} of creativity, rather than on creativity per se, which would alter the creative process and product \citep{jordanous2016four}. To this end, we promote that it is essential to encapsulate the system and its perceptible qualities from the participants while keeping the creative behaviour constant between the compared systems. Isolating these effects may offer us insight into how a system's embodiment can be changed for it to \textit{appear} more creative, even though the underlying creative processes remain the same.

We next introduce our experimental study framework, 
justify the controlled and independent variables, discuss their possible values and provide recommendations for the measurement of creativity as the dependent variable. We finish with conclusions and future work.

\section{Experimental Framework}

\begin{figure}
    \centering
    \includegraphics[width=0.90
\textwidth]{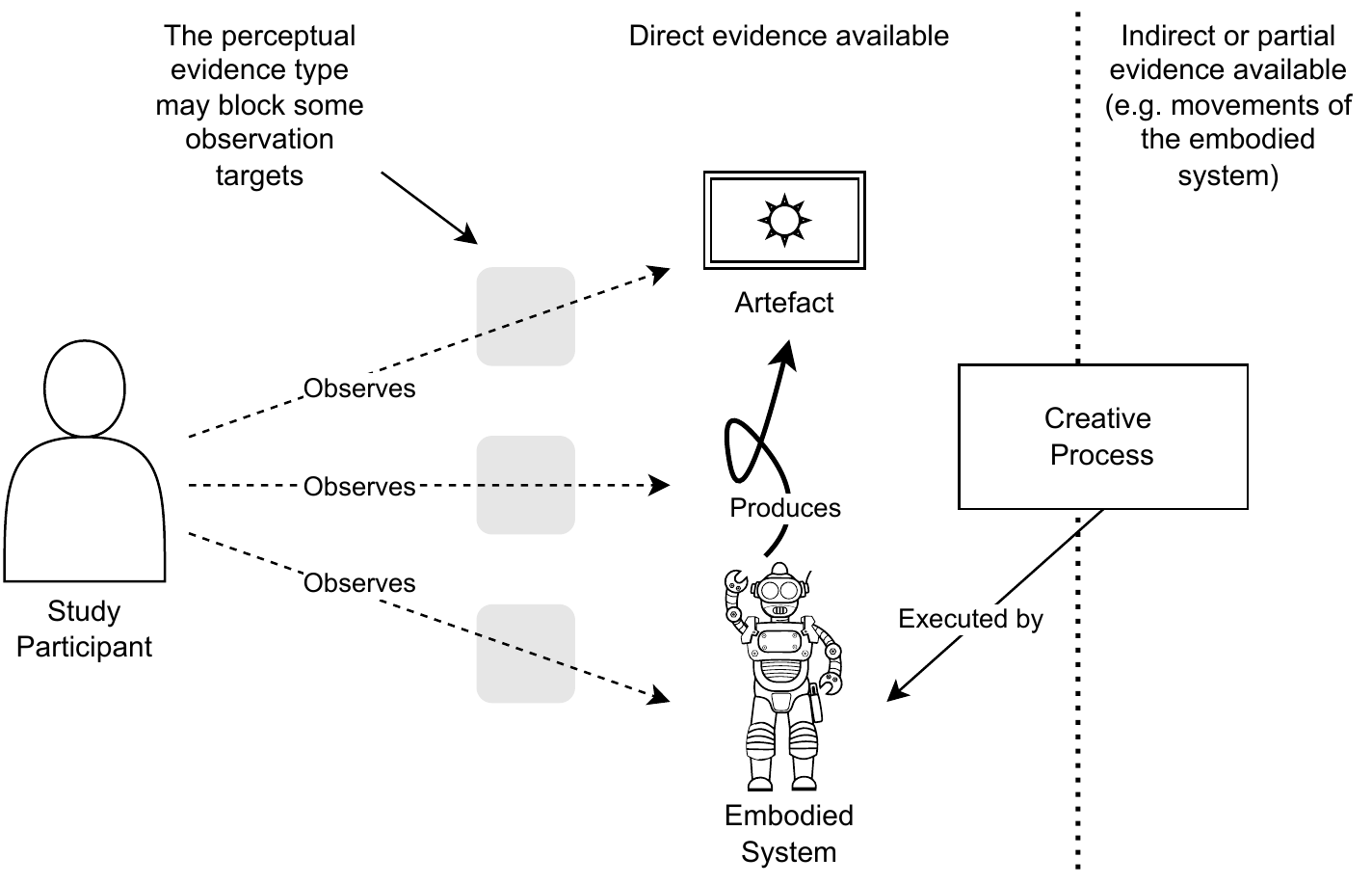}
    \caption{The general composition of the experiment setting. Depending on the perceptual evidence type designated for the participant, they may observe the artefact, the production of the artefact and/or the system producing the artefact. Moreover, the creative process may be partially hidden from the study participant regardless of the designated perceptual evidence type.}
    \label{fig:study-design}
\end{figure}

At the basis of our experimental framework, we factorise what is commonly denoted the \enquote{perception of creativity} \citep{colton2008creativity}
into (1) perceptual evidence, constraining the observable aspects of the system and its interaction, and (2) the assessment of the embodied system's creativity based on the available perceptual evidence. This perceptual evidence, in turn, is shaped by (3) the system's embodiment, and (4) the creative process which the system executes. Some aspects of this process typically remain hidden from the observer. We consequently model two independent, one dependent and one controlled variable. We manipulate the compared \emph{system's embodiment} (independent variable, IV) and which aspects of the systems' behaviour the participants can perceive, i.e. their \emph{perceptual evidence} (IV). We measure the participants' \emph{perceived creativity assessment} (dependent variable, DV), and hold the underlying \emph{creative process}, which is reflected in the perceptual evidence, constant between conditions (controlled variable, CV). Figure~\ref{fig:study-design} illustrates the general composition of the experiment setting.

We distinguish types of perceptual evidence based on three key concepts: the system, the process and the artefact. These correspond to three of the \enquote{four perspectives on creativity}, identified by \citet{rhodes1961analysis}, \citet{mooney1963conceptual} and potentially others \citep[cf.][]{jordanous2016four}. The system corresponds to the creative \emph{Person} \cite{rhodes1961analysis,mooney1963conceptual} or \emph{Producer} \citep{jordanous2016four}, e.g.~a plotter or a humanoid robot. Moreover, we can consider the creative \emph{Process} as the steps the \emph{Producer} engages in when being creative, such as producing movements with a pen on paper. Finally, we can consider the creative \emph{Product}, or artefact, as the outcome of this process, e.g. the finished drawing. We omit in our analysis the perspective of the \emph{Press}, which speaks to the socio-cultural environment or context of creativity. We consider this an important view as well, but it requires a further division to provide meaningful, manipulable variables for testing, which will be subject of future work.

The experimental framework imposes some restrictions on the studied systems. Namely, the creative process of the system must be kept the same while altering the embodiment or the perceptual evidence. Moreover, altering the perceptual evidence 
may be impossible in domains where the embodiment itself is part of the observed artefact, such as in dance. 

Next, we discuss each variable and its possible values in detail, and conclude the framework description with instructions for instantiating and executing a specific experiment.

\subsection{System Embodiment (IV)}
Our framework allows for comparing people's perceptions of multiple systems, each with a different embodiment. This independent variable defines the embodiments to be compared, with each value corresponding to one system and embodiment. In practice, several values could be represented by the same system through modifications to its embodiment. However, some embodiments are too different to be expressed on the same hardware and implementing very varied embodiments on different systems can enable comparisons with stronger effects.

While \enquote{embodied} is often considered synonymous with \enquote{robotic} 
systems, cognitive science has distinguished several more types of embodiment, which lend themselves naturally as values of our independent variable. We adopt the typology by \citet{ziemke2003s}, which has been slightly extended by \citet[][highlighted in \emph{italics}]{survey2021embodiment}:

\begin{description}
\item{\textbf{structural coupling}}, characterising systems that can perturbate, and, vice versa, be perturbated by their surrounding environment \citep{varela1991embodied}. 
\item{\textbf{historical}, characterising systems whose present state is the result of a history of structural couplings, developed through interactions with the environment over time  \citep{varela1991embodied, ziemke1999rethinking}.} 
\item{\textbf{virtual}}, \emph{characterising simulated systems embedded in and distinguished from a simulated environment. The virtual body can affect the environment and vice versa.}
\item{\textbf{physical}}, characterising systems with a physical body \citep{brooks1990elephants, pfeifer2001understanding} that can interact with the environment by being subjected to and exercising physical force. Includes most robots. 
\item{\textbf{organismoid}}, characterising \emph{virtually} or physically embodied systems with the same or similar form and sensorimotor equipment as living organisms. Approximations of a humanoid embodiment can be considered a subset.
\item{\textbf{organismic}}, applying to living \emph{and artificial} systems capable of autopoiesis, i.e. of maintaining their internal organisation and surrounding boundary against internal and external perturbations by means of self-producing processes \citep{von1920theoretische, maturana1987tree}.
\end{description}
While structural coupling establishes the baseline for what can be considered an embodied system, historical embodiment practically characterises any non-theoretical system. Consequently, we suggest only considering the latter four types and their sub-types for experimentation. A specific system can instantiate several types, which can be combined as values of the independent variable. For instance, one might be interested in comparing a system with virtual+humanoid (a subset of organismoid) with a system implementing physical+humanoid embodiment. 

While these types provide some initial orientation, they would typically be differentiated further. For instance, robots in the shape and with similar sensorimotor equipment to an ant, a dolphin, or a bird all share the same type of physical, organismoid but non-humanoid embodiment. While such diverse comparisons are possible, experimenters might also be interested in measuring the effect of subtle changes to the same embodiment, e.g.~equipping a humanoid, physically embodied system with a different hand.

The choice of values in a specific experiment will likely be informed by the overarching research question and the systems at hand. If no requirements are imposed, we argue that the most value to the community at present can be created by comparing those types of embodiment that most strongly differentiate existing CC studies: \emph{virtual} vs \emph{physical} embodiment, followed by humanoid vs non-humanoid, organismoid embodiment \citep{survey2021embodiment}.

\subsection{Perceptual Evidence (IV)}

We hold that how a person observes an embodied system, yielding different kinds of \textit{perceptual evidence} (PE), plays a key role in their attribution of creativity. This is why our framework factorises the perception of creativity into the creativity assessment and the \textit{perceptual evidence} (PE), and manipulates the latter as independent variable. We propose three types of perceptual evidence: the finished artefact (PE1), the artefact as it is being produced (PE2), and the embodied system as it is producing the artefact (PE3). In all cases, the participant is assumed to remain in a fixed location. These types as values of the independent variable can be further differentiated, e.g. by distinguishing different observation angles, or allowing the participant to observe freely. We next discuss how each of the proposed values might affect the attribution of creativity as dependent variable, using the \emph{four perspectives on creativity} framework~\citep{rhodes1961analysis,mooney1963conceptual,jordanous2016four}. 

\paragraph{PE1: The artefact}The participant only observes the finished, produced artefact, but not the creation process or the creating, embodied system. While the artefact evidences an embodied system's historical, sensorimotor interaction with its environment, the participant can only speculate on the details of its embodiment. This can serve as the baseline to which other types of perceptual evidence are compared.\\
\textbf{Creativity:} Observing only the artefact allows assessing creativity from the \emph{Product} perspective. While not observing the production makes it challenging to assess creativity from the \emph{Process} perspective, some hints of the \emph{Process} or \emph{Producer} qualities (e.g.~skill, imagination) may be evidenced in the artefact, becoming available for observation.


\paragraph{PE2: The process (incl.~PE1)}The participant observes the creation process of the artefact, e.g.~lines drawn on the screen, and the finished artefact (PE1). The embodiment of the system, however, is not explicitly revealed. This is straightforward when dealing with modifications of the same embodiment type, but comparisons between different types, e.g.~virtual and physical embodiment, require more sophisticated setups. While the creative process is controlled and should thus not reveal a particular embodiment, we yet hold that observing it might allow the participant to infer a particular kind of embodiment. Including PE2 accounts for such effects.\\
\textbf{Creativity:} Allowing the participant to observe the artefact as it is being produced enables (in suitable domains) assessing creativity from the \emph{Process} perspective. However, the full (creative) process of the system is not observable as the embodiment is left out. Using PE2 as an intermediate step between PE1 and PE3 allows comparing, e.g. how observing the artefact creation process itself affects the assessment of the produced artefacts (comparing PE1 and PE2).

\paragraph{PE3: The system (incl.~PE1, PE2)}The participant observes the embodied system while it is producing the artefact. In contrast to PE2, the participant can also observe how the embodied system actually operates, e.g.~drawing lines on a canvas. This allows for including all relevant embodiment types in the participant's observations, although organismic embodiment might be hard to realise within a sensible experimental time frame.\\
\textbf{Creativity:} This type allows the participant to assess a system's externally perceivable \emph{Process}, and observing the embodiment may give hints about the system's properties from the \emph{Producer} perspective. Comparing the same system on PE2 and PE3 enables studying the effect of observing the embodiment itself on the perceived creativity of the \emph{Products} and their production \emph{Process}.


\subsection{Perceived Creativity Assessment (DV)}
\label{ssec:assessing_perceived_creativity}

Depending on the goals of the experiment, a participant's perceived creativity assessment can be measured either qualitatively or quantitatively. While a comprehensive overview of creativity measurement practices is out of scope for this paper, we make the general recommendation to use standardised and CC-specific assessment practices such as SPECS~\citep{Jordanous2012SPECS}. Here, the study designer first specifies what is meant by creativity and then defines the criteria for how it can be assessed based on perceptual evidence. Moreover, combining observations (by the experimenter) and self-reporting of the participants has been known to produce apt results in studies assessing human experience while using creative systems~\citep{kantosalo2019quantifying_co-creative}. 

If conducting an interview, close attention must be paid to the formulation of the questions in order to foster insights on a particular aspect of the experiment, e.g.~the observed artefact, its creation process, or the system. Moreover, any question must explicitly distinguish between the observed evidence and unobserved variables: a~participant exposed to PE1 can be asked about the embodied system's properties, but it should be clearly communicated that answering the question requires them to imagine system properties which they have not directly observed.

\subsection{Creative Process (CV)}
In order to measure the effect of embodiment on the perception of creativity rather than on creativity per se, we must keep the creative process constant between conditions. We distinguish two means to accomplish this, with individual benefits and disadvantages: 
\begin{description}
\item[Online] The system must be modified ahead of the experiment to produce equal action sequences between runs, e.g.~by fixing random seeds, setting learning rates to zero, etc.~This comes with the advantage that the system can still condition its actions on previous sensor states in a \emph{closed loop} fashion, potentially producing a process which incorporates its present environment. However, it also risks variations between runs through e.g.~different participants prompting different reactions from the system. Moreover, it is easy to overlook any necessary modifications to keep the process constant. 
\item[Offline] All process-constituting actions are pre-recorded from a normal system run, and replayed for all conditions. This guarantees that the process is held constant between runs. However, it also implies that the system will stop reacting to any sensor inputs such as changes in its surroundings, including the participant. We hold that this change to \emph{open-loop} control might affect the human creativity assessment unnaturally. 
\end{description}

Even when keeping the system's internal action signals constant, maintaining the outside process as perceived by the participant the same may not always be possible. In particular, in physically embodied systems, actuator and sensor noise and/or fidelity will yield slight variations in the observable process, and, thus, variations in the resulting artefact. How sensitive the measurement of the dependent variable is to such fluctuations is an open research question. We recommend accounting for this through \emph{repeated measurements} of the same condition and ensuring that the variations only produce minuscule differences in the produced artefacts.

Lastly, we anticipate that the human assessment of creativity might be affected by many latent properties of the creative process, e.g. its length, execution speed, or complexity. To account for such biases, we can evaluate multiple creative processes with varying properties, effectively turning the creative process into an independent variable.


\subsection{Instantiating an Experiment}

To instantiate an experiment with this framework for a specific research question, one must make the following decisions:


\begin{itemize}
    \item Which of the independent variables to manipulate (system embodiment, perceptual evidence, both). If both independent variables are manipulated, we have a \emph{factorial design}.
    \item Which values to assume for each of these independent variables. To yield more than one condition for a comparative measurement with only one independent variable, it is crucial to select at least two or more different values of perceptual evidence, or embodiments.
    \item Which creative process to perform and control for.
    \item How to measure the assessment of creativity as dependent variable? 
    \item Whether to expose each participant to one, all, or only some conditions, yielding a \emph{between/within/mixed}  experimental design, respectively.
    \item How to analyse the collected data (e.g. exploratory statistics and statistical significance testing for quantitative, thematic analysis for qualitative data).
\end{itemize}

By controlling the perceptual evidence, we can assess how strongly the subjective assessment of embodied CC depends on what data on the artefact-process-system whole has been gathered. This can also be interpreted as manipulating not the system's, but the participant's sensorimotor embodiment, which serves as additional justification for including perceptual evidence in our framework which focuses on embodiment effects. The experiment also allows studying how embodiment affects the perceived creativity by altering embodiment (e.g.~virtual and physical) while keeping the process and the perceptual evidence the same. Both effects can be investigated together in a factorial design with appropriate analysis methods. The choice of experimental design (between/within/mixed) can for instance be informed by the number of conditions, e.g. to prevent participant fatigue, and the available hardware (e.g.~only one robot which must be repeatedly modified to create various embodiment study conditions). Crucially, if the same participant is exposed to multiple conditions, potential order effects must be mitigated by randomising the exposition order for each participant.

\subsection{Experiment Procedure}
The experiment should be executed for individuals, and not groups of participants, in a lab setting and without distractions. After (1) receiving an introduction to the experiment, the participant must (2) provide consent to the use of their data and (3) provide demographic information. The experimenter should also (4) rule out any risk of bias in conversation with the participant, e.g.~due to existing knowledge of the system or its designer(s). The participant is then (5) exposed to perceptual evidence of a single embodied system (i.e.~process and embodiment). Data is recorded during or after this exposure, depending on the selected measurement. The experiment can stop (between), commence with all remaining conditions (within) or cover a subset (mixed), depending on the choice of experimental design. The experiment can be concluded with a (6) post-study questionnaire and (7) debriefing session. 

\section{Conclusions and Future Work}

We have presented an experimental framework for studying how a computational creativity (CC) system's embodiment affects the human perception of its creativity. To this end, we have factorised the perception of creativity into a person's \emph{creativity assessment} (dependent variable), and \emph{perceptual evidence} (independent variable) of the system and its \emph{embodiment} (independent variable) as a mediator. Moreover, we isolated the impact of embodiment on the perception of creativity from its impact on creativity per se by keeping the artefact production \emph{process} (controlled variable) constant between conditions. We distinguished six types of embodiment based on existing typologies \citep{ziemke2003s,survey2021embodiment}, discussed two distinct means to control the creative process, and put forward three types of perceptual evidence, corresponding to the \emph{Product}, \emph{Process} and \emph{Producer} in the \emph{four perspectives on creativity} framework \citep{rhodes1961analysis,mooney1963conceptual,jordanous2016four}. Using the experimental framework, we can answer many interesting questions about the perceived creativity of an embodied system, e.g. \enquote{How does observing the artefact production process affect the creativity assessment when compared to observing only the finished artefact?}, or \enquote{How does a more human-like embodiment impact the perceived creativity assessment?}.

The core framework presented in this paper can be further extended. For example, we have focused on visual observations, but allowing other or multi-modal observations is feasible and worthwhile. 
Moreover, we could treat other types of evidence, such as framing \citep{charnley2012notion} -- e.g.~a piece of text covering the motivation behind the artefact and other not directly observable information about the system -- as a controlled or independent variable. Additionally, further aspects from the socio-cultural environment of the creative phenomenon -- i.e. the \emph{Press} in the \emph{four perspectives on creativity} framework~\cite{jordanous2016four,rhodes1961analysis} -- could be brought in for analysis.

Our immediate next step is to instantiate this framework and perform a first principled study on the effect of embodiment and perceptual evidence on the human perception of computational creativity. Our focus will be on comparing the effect of different virtual and physical embodiments on the assessment of creativity, and on how different kinds of perceptual evidence moderate this effect. Another goal for future work is to use the framework on an existing, established embodied CC system to allow assessing its perceived creativity either with different embodiment modifications or different perceptual evidence. Finally, we would like to test the flexibility of this framework by conducting an experiment outside the lab, e.g. in an exhibition space, in order to improve the ecological validity of our findings. {We consider this framework an important step toward better embodied CC research \citep{survey2021embodiment}, enabling empirical studies to inform the design and comparison of CC systems that interact with people.

\subsection{Acknowledgments}
SL and AK were funded by the Academy of Finland (project \#328729), and CG at an early stage by the Academy of Finland programme “Finnish Center for Artificial Intelligence” (FCAI).

\bibliography{iccc}

\begin{thebibliography}{20}
\expandafter\ifx\csname natexlab\endcsname\relax\def\natexlab#1{#1}\fi
\providecommand{\url}[1]{\texttt{#1}}
\providecommand{\href}[2]{#2}
\providecommand{\path}[1]{#1}
\providecommand{\DOIprefix}{doi:}
\providecommand{\ArXivprefix}{arXiv:}
\providecommand{\URLprefix}{URL: }
\providecommand{\Pubmedprefix}{pmid:}
\providecommand{\doi}[1]{\href{http://dx.doi.org/#1}{\path{#1}}}
\providecommand{\Pubmed}[1]{\href{pmid:#1}{\path{#1}}}
\providecommand{\bibinfo}[2]{#2}
\ifx\xfnm\relax \def\xfnm[#1]{\unskip,\space#1}\fi
\bibitem[{Colton(2008)}]{colton2008creative_tripod}
\bibinfo{author}{S.~Colton},
\newblock \bibinfo{title}{Creativity versus the perception of creativity in
  computational systems},
\newblock in: \bibinfo{booktitle}{AAAI Spring Symposium: Creative Intelligent
  Systems}, \bibinfo{year}{2008}, pp. \bibinfo{pages}{14--20}.
\bibitem[{Lamb et~al.(2015)Lamb, Brown, and Clarke}]{lamb2015}
\bibinfo{author}{C.~Lamb}, \bibinfo{author}{D.~G. Brown},
  \bibinfo{author}{C.~Clarke},
\newblock \bibinfo{title}{{Human Competence in Creativity Evaluation}},
\newblock in: \bibinfo{booktitle}{Proc. ICCC}, \bibinfo{year}{2015}, pp.
  \bibinfo{pages}{102--109}.
\bibitem[{Mumford and Ventura(2015)}]{mumford15}
\bibinfo{author}{M.~Mumford}, \bibinfo{author}{D.~Ventura},
\newblock \bibinfo{title}{The man behind the curtain: {O}vercoming skepticism
  about creative computing},
\newblock in: \bibinfo{booktitle}{Proc. ICCC}, \bibinfo{year}{2015}, pp.
  \bibinfo{pages}{1--7}.
\bibitem[{Guckelsberger et~al.(2017)Guckelsberger, Salge, and
  Colton}]{guckelsberger2017addressing}
\bibinfo{author}{C.~Guckelsberger}, \bibinfo{author}{C.~Salge},
  \bibinfo{author}{S.~Colton},
\newblock \bibinfo{title}{{Addressing the ``Why?" in Computational Creativity:
  A Non-Anthropocentric, Minimal Model of Intentional Creative Agency}},
\newblock in: \bibinfo{booktitle}{{Proc. ICCC}}, \bibinfo{year}{2017}, pp.
  \bibinfo{pages}{128--135}.
\bibitem[{Charnley et~al.(2012)Charnley, Pease, and
  Colton}]{charnley2012notion}
\bibinfo{author}{J.~W. Charnley}, \bibinfo{author}{A.~Pease},
  \bibinfo{author}{S.~Colton},
\newblock \bibinfo{title}{{On the Notion of Framing in Computational
  Creativity}},
\newblock in: \bibinfo{booktitle}{{Proc. ICCC}}, \bibinfo{year}{2012}, pp.
  \bibinfo{pages}{77--81}.
\bibitem[{Colton et~al.(2020)Colton, Pease, Guckelsberger, McCormack, and
  Llano}]{Colton2020}
\bibinfo{author}{S.~Colton}, \bibinfo{author}{A.~Pease},
  \bibinfo{author}{C.~Guckelsberger}, \bibinfo{author}{J.~McCormack},
  \bibinfo{author}{M.~T. Llano},
\newblock \bibinfo{title}{{On the Machine Condition}},
\newblock in: \bibinfo{booktitle}{{Proc. ICCC}}, \bibinfo{year}{2020}, pp.
  \bibinfo{pages}{342--349}.
\bibitem[{Guckelsberger et~al.(2021)Guckelsberger, Kantosalo,
  Negrete-Yankelevich, Takala et~al.}]{survey2021embodiment}
\bibinfo{author}{C.~Guckelsberger}, \bibinfo{author}{A.~Kantosalo},
  \bibinfo{author}{S.~Negrete-Yankelevich}, \bibinfo{author}{T.~Takala},
  et~al.,
\newblock \bibinfo{title}{{Embodiment and Computational Creativity}},
\newblock in: \bibinfo{booktitle}{Proc. ICCC}, \bibinfo{year}{2021}, pp.
  \bibinfo{pages}{192--201}.
\bibitem[{Ziemke(2003)}]{ziemke2003s}
\bibinfo{author}{T.~Ziemke},
\newblock \bibinfo{title}{{What’s That Thing Called Embodiment?}},
\newblock in: \bibinfo{booktitle}{Proc. Annual Meeting of the Cognitive Science
  Society}, volume~\bibinfo{volume}{25}, \bibinfo{year}{2003}, pp.
  \bibinfo{pages}{1305--1310}.
\bibitem[{Jordanous(2016)}]{jordanous2016four}
\bibinfo{author}{A.~Jordanous},
\newblock \bibinfo{title}{{Four PPPPerspectives on Computational Creativity in
  Theory and in Practice}},
\newblock \bibinfo{journal}{Connection Science} \bibinfo{volume}{28}
  (\bibinfo{year}{2016}) \bibinfo{pages}{194--216}.
\bibitem[{Colton(2008)}]{colton2008creativity}
\bibinfo{author}{S.~Colton},
\newblock \bibinfo{title}{Creativity versus the perception of creativity in
  computational systems.},
\newblock in: \bibinfo{booktitle}{AAAI spring symposium: creative intelligent
  systems}, volume~\bibinfo{volume}{8}, \bibinfo{organization}{Palo Alto, CA},
  \bibinfo{year}{2008}, p.~\bibinfo{pages}{7}.
\bibitem[{Rhodes(1961)}]{rhodes1961analysis}
\bibinfo{author}{M.~Rhodes},
\newblock \bibinfo{title}{An analysis of creativity},
\newblock \bibinfo{journal}{The Phi delta kappan} \bibinfo{volume}{42}
  (\bibinfo{year}{1961}) \bibinfo{pages}{305--310}.
\bibitem[{Mooney(1963)}]{mooney1963conceptual}
\bibinfo{author}{R.~Mooney},
\newblock \bibinfo{title}{{A Conceptual Model for Integrating Four Approaches
  to the Identification of Creative Talent}},
\newblock in: \bibinfo{editor}{C.~W. Taylor}, \bibinfo{editor}{F.~Barron}
  (Eds.), \bibinfo{booktitle}{{Scientific Creativity: Its Recognition and
  Development}}, \bibinfo{publisher}{Wiley}, \bibinfo{year}{1963}, pp.
  \bibinfo{pages}{331--340}.
\bibitem[{Varela et~al.(1991)Varela, Rosch, and Thompson}]{varela1991embodied}
\bibinfo{author}{F.~J. Varela}, \bibinfo{author}{E.~Rosch},
  \bibinfo{author}{E.~Thompson}, \bibinfo{title}{{The Embodied Mind}},
  \bibinfo{publisher}{MIT Press}, \bibinfo{year}{1991}.
\bibitem[{Ziemke(1999)}]{ziemke1999rethinking}
\bibinfo{author}{T.~Ziemke},
\newblock \bibinfo{title}{{Rethinking Grounding}},
\newblock in: \bibinfo{booktitle}{{Understanding Representation in the
  Cognitive Sciences}}, \bibinfo{publisher}{Springer}, \bibinfo{year}{1999},
  pp. \bibinfo{pages}{177--190}.
\bibitem[{Brooks(1990)}]{brooks1990elephants}
\bibinfo{author}{R.~A. Brooks},
\newblock \bibinfo{title}{{Elephants Don't Play Chess}},
\newblock \bibinfo{journal}{Robotics and Autonomous Sys.} \bibinfo{volume}{6}
  (\bibinfo{year}{1990}) \bibinfo{pages}{3--15}.
\bibitem[{Pfeifer and Scheier(2001)}]{pfeifer2001understanding}
\bibinfo{author}{R.~Pfeifer}, \bibinfo{author}{C.~Scheier},
  \bibinfo{title}{{Understanding Intelligence}}, \bibinfo{publisher}{MIT
  Press}, \bibinfo{year}{2001}.
\bibitem[{Von~Uexk{\"u}ll(1920)}]{von1920theoretische}
\bibinfo{author}{J.~Von~Uexk{\"u}ll}, \bibinfo{title}{{Theoretische Biologie}},
  \bibinfo{publisher}{Paetel}, \bibinfo{year}{1920}.
\bibitem[{Maturana and Varela(1987)}]{maturana1987tree}
\bibinfo{author}{H.~R. Maturana}, \bibinfo{author}{F.~J. Varela},
  \bibinfo{title}{{The Tree of Knowledge: The Biological Roots of Human
  Understanding}}, \bibinfo{publisher}{New Science Library/Shambhala
  Publications}, \bibinfo{year}{1987}.
\bibitem[{Jordanous(2012)}]{Jordanous2012SPECS}
\bibinfo{author}{A.~Jordanous},
\newblock \bibinfo{title}{{A Standardised Procedure for Evaluating Creative
  Systems: Computational Creativity Evaluation Based on What it is to be
  Creative}},
\newblock \bibinfo{journal}{Cognitive Comp.} \bibinfo{volume}{4}
  (\bibinfo{year}{2012}) \bibinfo{pages}{246--279}.
\bibitem[{Kantosalo and Riihiaho(2019)}]{kantosalo2019quantifying_co-creative}
\bibinfo{author}{A.~Kantosalo}, \bibinfo{author}{S.~Riihiaho},
\newblock \bibinfo{title}{{Quantifying Co-Creative Writing Experiences}},
\newblock \bibinfo{journal}{Digital Creativity} \bibinfo{volume}{30}
  (\bibinfo{year}{2019}) \bibinfo{pages}{23--38}.

\end{thebibliography}

\end{document}